\documentclass[showpacs,amsmath,amssymb,twocolumn,pra,superscriptaddress,notitlepage]{revtex4-1}

\usepackage{subfigure}
\usepackage{amsfonts}
\usepackage{amssymb}
\usepackage{amscd}
\usepackage{color}
\usepackage{amsmath}    % need for subequations
\usepackage{multirow}
\usepackage{graphicx}% Include figure files
\usepackage{dcolumn}% Align table columns on decimal point
\usepackage{bm}% bold math
\newcommand{\bra}[1]{\mbox{$\left\langle #1 \right|$}}
\newcommand{\ket}[1]{\mbox{$\left| #1 \right\rangle$}}

\begin{document}
\preprint{APS/123-QED}
%\title{Coherence consumption in homodyne detection based quantum random number generators}
%\title{Fundamental limits of quantum random number generation based on coherence measures}
\title{Randomness quantification of coherent detection}

\author{Hongyi Zhou}
\affiliation{Center for Quantum Information, Institute for Interdisciplinary Information Sciences, Tsinghua University, Beijing, 100084 China}
%\email{yuanxiao12@mails.tsinghua.edu.cn}
\author{Pei Zeng}
\affiliation{Center for Quantum Information, Institute for Interdisciplinary Information Sciences, Tsinghua University, Beijing, 100084 China}
\author{Mohsen Razavi}
\affiliation{School of Electronic and Electrical Engineering, University of Leeds, Leeds, LS2 9JT, UK}
\author{Xiongfeng Ma}
\affiliation{Center for Quantum Information, Institute for Interdisciplinary Information Sciences, Tsinghua University, Beijing, 100084 China}

%%%%%%%%%%%%%%%%%%%%%%%%%%%%%%%%%%%%%%%%%%
\begin{abstract}
Continuous-variable quantum cryptographic systems, including random number generation and key distribution, are often based on coherent detection. The essence of the security analysis lies in the randomness quantification. Previous analyses employ a semi-quantum picture, where the strong local oscillator limit is assumed. Here, we investigate the randomness of homodyne detection in a full quantum scenario by accounting for the shot noise in the local oscillator, which requires us to develop randomness measures in the infinite-dimensional scenario. Similar to the finite-dimensional case, our introduced measure of randomness corresponds to the relative entropy of coherence defined for an infinite-dimensional system. Our results are applicable to general coherent detection systems, in which the local oscillator is inevitably of finite power. As an application example, we employ the analysis method to a practical vacuum-fluctuation quantum random number generator and explore the limits of generation rate given a continuous-wave laser.

%True randomness of a system, guaranteed by quantum mechanics, can be quantified with its coherence on the computational basis. Coherence as well as randomness quantification have been well defined for finite dimensional discrete variables and it is interesting to extend them to infinite dimensional systems such as coherent detection. In this work, we investigate the randomness generation in a shot-noise QRNG scheme based on homodyne detection, which is limited by the infinite dimensional coherence. For practical systems, we give an upper and lower bound of random number generation rate. Such randomness quantification technique would be very helpful in the security analysis of continuous-variable quantum key distribution. %In this paper, we analyse the quantum randomness origin in a vacuum QRNG from the first principle and explore the limit of random number generation rate given a CW laser.
\end{abstract}



%\submitto{\NJP}

\maketitle
\section{Introduction}
%As one of the most practical fields in quantum information science, quantum cryptography has addressed two major problems: how to securely exchange secret keys, and how to verifiably generate random numbers.

Quantum cryptography, the most practical field in quantum information science, has two major tasks --- key distribution and randomness generation. Quantum key distribution (QKD) allows communication partners to share private keys in the presence of an eavesdropper, Eve, whose power is only limited by quantum mechanics \cite{Bennett1984Quantum,Ekert1991Quantum}. Quantum random number generation (QRNG) aims at providing unpredictable random numbers \cite{MaQRNG,RevModPhys.89.015004}. The main theoretical focus of both cryptographic tasks lies in the security analysis, which ensures that Eve cannot predict the key or random numbers. Mathematically, the definitions of privacy in the key bits and unpredictability in the random numbers are the same. Thus, it is expected that security analyses in QKD can also be applied to QRNG and vice versa.

%xma: Shall we say here that we will focus on QRNG below, but similar results should be directly applied to QKD.

%another way, discuss the development in both tasks first, including satellite, and then move on...

There are mainly two categories of schemes for quantum cryptographic systems, namely, discrete variable and continuous variable. Continuous-variable cryptography \cite{GG02,grosshans2003quantum} employs Gaussian modulation and coherent detection, e.g., homodyne detection and heterodyne detection. 
%Compared with discrete-variable cryptography, it enhances the capacity of information carriers, which will result in higher key rates or randomness. Meanwhile,
These are standard techniques in classical telecommunications, which could make continuous-variable optical components robust and economic. From the theoretical point of view, it is crucial to study the mechanism of coherent detection for the security analysis of continuous-variable cryptography. Without loss of generality, we will focus on continuous-variable QRNG systems below. Similar results should also be applicable to QKD systems.

Continuous-variable QRNG schemes \cite{Gabriel10,STZ10,Symul11,Jofre11,SCK16,PhysRevLett.118.060503,xu2017high,haylock2018multiplexed,guo2018enhancing,avesani2018secure,zheng20186,furst2010high,yan2014multi,Qi10,Xu12,Nie14,Nie15,Nie16,yang2016oe,sun17pra} offer some advantage over conventional discrete-variable ones \cite{jennewein2000fast,stefanov2000optical,Ma15,Ma16} in both performance and practicality, especially the ones exploiting quadrature fluctuations of optical fields \cite{Gabriel10,STZ10,Symul11,Jofre11,SCK16,PhysRevLett.118.060503,xu2017high,haylock2018multiplexed,guo2018enhancing,avesani2018secure,zheng20186} or laser phase fluctuations \cite{Qi10,Xu12,Nie14,Nie15,Nie16,yang2016oe,sun17pra}, pushing the generation rate from Mbps to the Gbps regime. The substantial improvement in randomness generation performance is mainly attributed to the coherent detection technique, which replaces single-photon detectors with high-performance photodetectors, gets rid of the restriction of detector dead time, and yields a higher sampling rate.

For these continuous-variable QRNG schemes based on coherent detection, a physical model from the first principle along with rigorous randomness quantification is still missing. Former models of coherent detection QRNGs assumed that the local oscillator in use behaves classically \cite{Gabriel10,STZ10,Symul11,Jofre11,SCK16,PhysRevLett.118.060503,xu2017high,haylock2018multiplexed,guo2018enhancing,avesani2018secure,zheng20186,Ma13,Zhou15}. In that case, by controlling the phase of the local oscillator $\phi$, different quadratures $\hat{x}(\phi)=1/2(\hat{a}e^{-i\phi}+\hat{a}^{\dag}e^{i\phi})$ of the incoming mode of light, with annihilation operator $\hat a$, can then be measured. This leads to a continuum of measurement outcomes implying that the amount of randomness extracted from single round of detection is divergent with high detection resolution, which is rather counter-intuitive. Another issue lies in randomness quantification, where conventional approaches are based on classical min-entropy function \cite{Gabriel10,STZ10,Qi10,Xu12,Nie15,Nie16}. Such quantifiers may suffer from side information in the measurement outcomes, i.e., the quantum state before measurement may be entangled with some ancillary systems held by the adversary. Though the nominal output randomness can be calculated by the measurement statistics, the intrinsic randomness that comes from quantum measurement stays unknown.

%But, if we look more closely, what a homodyne receiver with the vacuum state as the input would effectively measure is the difference of photon counts in each of its legs driven by a coherent state that corresponds to the local oscillator split by a balanced beam splitter. In this case, we can argue that the randomness in the outcome is a result of the shot-noise effect in our photodetection. For that reason, we refer to this scheme by shot-noise driven QRNG.

In this work, we properly model the coherent detection and provide a rigorous analysis of the randomness origin, quantification, and fundamental limits. By modelling the local oscillator quantum mechanically with a pure coherent state, we can look more closely at the mechanism of the coherent detection. What a coherent detection would effectively measure is the photon number difference between different legs. In this case, we can argue that the randomness in the outcome is a result of the shot-noise effect in the photodetection. For that reason, we refer to the continuous-variable QRNG scheme with coherent detection by shot-noise driven QRNG, whose measurement outcomes form a {\em discrete}, rather than continuous, infinite-dimensional space.

Meanwhile, in order to accurately calculate the intrinsic randomness in such a QRNG, we apply the rigorous and powerful tool of quantum coherence \cite{baumgratz2014quantifying}, which has been related to quantum randomness in \cite{yuan2015intrinsic}. For instance, in the quantum information context, the $Z$-basis measurement on the qubit $(|0\rangle + |1\rangle)/\sqrt{2}$ would result in either basis states with equal probability. The result of such a measurement is unpredictable. A simple implementation of this idea is based on measuring the relative phase or polarization of a single photon \cite{stefanov2000optical}. One can get a similar result if, instead of a superposed state, a mixed state $(\ket{0}\bra{0}+\ket{1}\bra{1})/2$ is measured. In the latter case, however, we cannot rule out the possibility of the input state being entangled with another external system. In fact, we can purify our mixed state into a Bell state, in which case, an adversary party, who may hold the other part of the Bell state, can fully predict the outcome of the measurement. There is, in fact, no intrinsic (unpredictable) quantum randomness in this mixed-state case, and it only represents sheer classical randomness. The transition from fully random in the case of the superposition state to no quantum randomness for the mixed state indicates a correspondence between coherence of a state and how much quantum randomness can be extracted from it. In fact, it has been shown that, for finite-dimensional states, the relative entropy of coherence is an intrinsic randomness quantifier \cite{Yuan16}. In this paper, we extend this result to the infinite dimensional case and quantify the randomness in shot-noise driven QRNG with the help of infinite dimensional coherence \cite{zhang2016quantifying}. We believe such an analysis should be a standard approach for randomness quantification of QRNGs based on coherent detection, and be further widely employed in other continuous-variable cryptography systems.

The rest of this paper is organised as follows. In Sec.~\ref{Sec:physicalmodel}, we review the shot-noise driven QRNG structure and show that to properly quantify its generated randomness, we need to employ relevant measures for discrete infinite dimensional variables. Such measures are derived in Sec.~\ref{sec:limits} and their correspondence with infinite dimensional coherence on Fock basis is shown. We then quantify the randomness in shot-noise driven QRNGs and find practical rate bounds for its realistic implementations in Sec.~\ref{Sec:practicalbounds} before concluding the paper in Sec.~\ref{sec:conc}.

\section{Shot-Noise Driven QRNG}\label{Sec:main}
\subsection{Physical model of homodyne detection}\label{Sec:physicalmodel}

\begin{center}
\begin{figure*}[hbt]
\centering
\includegraphics[width=14 cm]{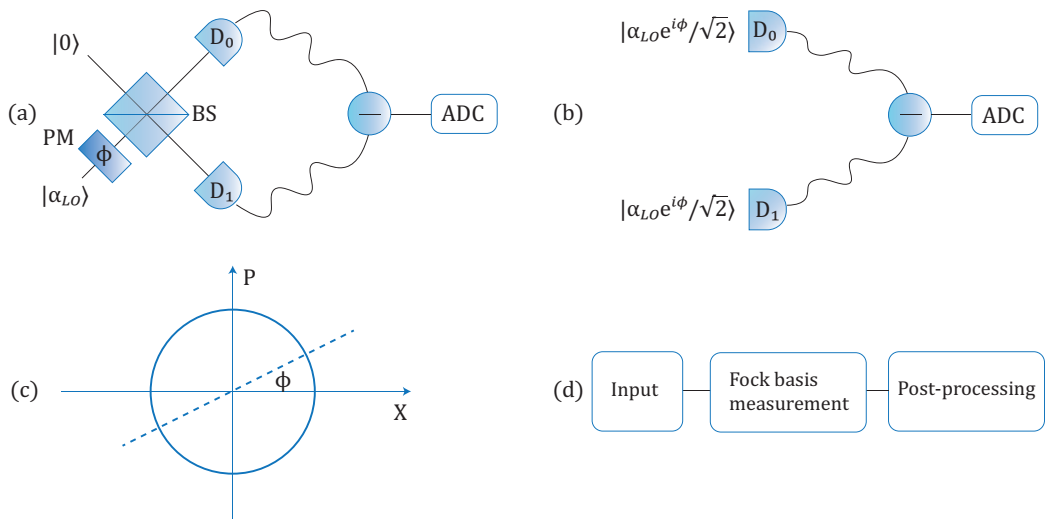}
\caption{(a) Schematic diagram of a shot-noise driven QRNG. A homodyne receiver measures a certain quadrature $x(\phi)$ of the vacuum state, which is controlled by the phase modulator. (b) Equivalent setting of (a). The two input coherent states $\ket{\alpha_{LO}e^{i\phi}/\sqrt{2}}$ have the same phase, but their intensities are independent. The output is proportional to the photon-number difference measured by the two detectors. Here, the randomness originates from the shot-noise effect. (c) Phase space presentation of classical modelled homodyne detection measuring $x(\phi)$ quadrature of a vacuum state. (d) Generalized flow chart of a shot-noise driven QRNG. The whole process can be divided into a quantum phase performing Fock basis measurement on certain input states and a classical phase performing a post-processing on the Fock basis measurement outcomes. LO: local oscillator; PM: phase modulator; BS: beam splitter; $D_{0,1}$: photo detector; ADC: analogue-to-digital converter. }%Alternatively, this device measures the photon-number difference detected by its two detectors. The randomness can then be associated to the shot-noise effect.}
\label{fig:settings}
\end{figure*}
\end{center}

Here we first focus on a shot-noise driven QRNG model which is based on homodyne detection of a vacuum state. A slightly modified version of this model can also be applied to other coherent detections, such as heterodyne detection.
% Note that all the shot-noise driven QRNGs behaves similarly, thus the randomness of other shot-noise driven QRNGs can be quantified in the same way.
% A shot-noise driven QRNG is based on homodyne detection of a vacuum state.
A schematic diagram of homodyne detection is shown in Fig.~\ref{fig:settings}(a). A local oscillator (LO) in coherent state $|\alpha_{\rm LO}\rangle$ is coupled to a vacuum state at a 50:50 beam splitter (BS). The two output modes are then measured by two identical photodetectors. The resulting currents are subtracted from each other and converted to bits by an analogue-to-digital converter (ADC).

%\begin{figure}[hbt]
%\centering
%\includegraphics[width=6 cm]{equivalentsettings}
%\caption{An equivalent setting for the scheme of Fig.~\ref{Fig:settings}. The two input coherent states $\ket{\alpha_{LO}/\sqrt{2}}$ are correlated in phase. The output is proportional to the photon-number difference measured by the two detectors. Here, the randomness originates from the shot-noise effect.} \label{equivalentsettings}
%\end{figure}

Such a process is expected to introduce random numbers. In previous analyses \cite{Gabriel10,Symul11,STZ10,SCK16}, the LO is modelled classically as a plane wave (in the limit of strong LOs). We show the detailed classical description in Appendix \ref{App:ClassicalModel}. This device practically measures $\hat x(\phi)$ quadrature of the vacuum state following Gaussian distribution, where $\phi$ is the modulated phase of the LO. In phase space, such a measurement is a cross-section of the Wigner function of the vacuum state (Fig.~\ref{fig:settings}(c)).

\begin{figure}[hbt]
\centering
\includegraphics[width=8.5 cm]{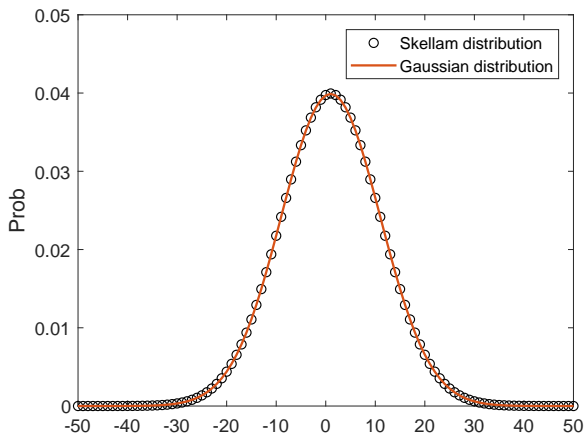}
\caption{Comparison of a Skellam distribution, given in Eq.~\eqref{eq:skellam}, and a Gaussian distribution, with the same mean $0$ and variance $2\mu=100$.}
\label{fig:ADCprob}
\end{figure}

Now, more precisely, we quantum mechanically characterize the LO as a pure coherent state
\begin{equation}
\ket{\alpha_{\rm LO}}=e^{-\frac{|\alpha_{\rm LO}|^2}{2}}\sum_n \frac{\alpha_{\rm LO}^n}{\sqrt{n!}} \ket{n},
\end{equation}
where $\alpha_{\rm LO}$ is a complex number and $\ket{n}$ is a Fock state with $n$ photons. Then we can model the module in Fig.~\ref{fig:settings}(a) by that of Fig.~\ref{fig:settings}(b). Each photodetector performs a Fock basis measurement on $\ket{\alpha_{\rm LO}e^{i\phi}/\sqrt{2}}$.
%, where we assume that the quantum efficiency of photodetectors has been absorbed into the LO amplitude. In that case, the photodetectors will have unity efficiencies, and, ignoring dark current effects, will measure the photon number in each of incoming pulses.
Because of the shot-noise effect, the output of both Fock basis measurements would follow a Poisson distribution
\begin{equation}\label{eq:poisson}
\begin{aligned}
p^{\mathcal{P}}_j(\mu)&=|\langle j|\alpha_{\rm LO}e^{i\phi}/\sqrt{2} \rangle|^2 \\
&=e^{-\mu}\frac{\mu^j}{j!},
\end{aligned}
\end{equation}
with a mean of $\mu = |\alpha_{\rm LO}|^2/2$ and independent of the modulated phase $\phi$. If we denote the measured photon number by detector $D_i$, $i=0,1$, by $N_i$, the input to the ADC would then be proportional to the photon number difference $N_d = N_0 - N_1$. It can be shown that $N_d$, as a difference of two independent Poisson distributions, follows Skellam distribution \cite{skellam1945frequency} given by
\begin{equation}\label{eq:skellam}
\begin{aligned}
p^{\mathcal{S}}_j(\mu) &=\Pr(N_d=j) \\
&=\left\{ \begin{aligned}
e^{-2\mu}I_j(2\mu)  & \quad j>0 \\
e^{-2\mu}I_{-j}(2\mu) &\quad   j<0
\end{aligned} \right.
\end{aligned}
\end{equation}
where $I_j(2\mu)$ is the modified Bessel function given by \cite{abramowitz1966handbook}
\begin{equation}
I_j(2\mu)=\sum_{m=0}^{\infty}\frac{1}{m!\Gamma(m+j+1)}{\mu}^{2m+j}.
\end{equation}
Figure \ref{fig:ADCprob} shows the Skellam distribution at $\mu = 50$. It can be seen that it has a symmetric form getting its maximum value at $j = 0$. For sufficiently large values of $\mu$, the Skellam distribution can be well approximated by a Gaussian distribution.

Our physical model of shot-noise driven QRNG can be easily generalized into coherent states input and general coherent detections (homodyne detection and heterodyne detection).
For example, in homodyne detection, an $X$ quadrature measurement on the coherent state input $\ket{\beta}$ leads to a Skellam distribution of
\begin{equation}
p_j(\mu_1,\mu_2)=e^{-(\mu_1+\mu_2)}\left(\frac{\mu_1}{\mu_2}\right)^{j/2}I_j(2\sqrt{\mu_1\mu_2}).
\end{equation}
where $\mu_1$ and $\mu_2$, where
\begin{equation}
\begin{aligned}
\mu_1&=\frac{|\beta+\alpha_{LO}|^2}{2} ,\\
\mu_2&=\frac{|\beta-\alpha_{LO}|^2}{2}.
\end{aligned}
\end{equation}
and if we replace homodyne detection with heterodyne detection, there are two output Skellam distributions $p_j(\mu_3,\mu_4)$ and $p_j(\mu_5,\mu_6)$ where
\begin{equation}
\begin{aligned}
\mu_{3,4}&=|\frac{\beta}{2}\pm \frac{\alpha_{LO1}}{\sqrt{2}}|^2 \\
\mu_{5,6}&=|\frac{\beta}{2}\pm \frac{\alpha_{LO2}}{\sqrt{2}}|^2
\end{aligned}
\end{equation}
Here $\ket{\alpha_{LO1}}$ and $\ket{\alpha_{LO2}}$ are local oscillators in heterodyne detection. For simplicity, we analyse the vacuum input and homodyne detection in the following text, but the methods for other cases are the same.

To fundamentally study the quantum randomness generated by the shot-noise driven QRNG, we have to separate classical sources of randomness from the underlying quantum phenomena. In our case, the electric noise of the receiver, for instance, would contribute to classically generated randomness and needs to be extracted out using distillation techniques. True intrinsic randomness comes from the photon number difference explained above. We then deal with an infinite dimensional, but discrete, random variable. In the next section, we derive a proper measure of randomness for such cases.

\subsection{Randomness origin and quantification: infinite dimensional coherence}
\label{sec:limits}
%\subsection{Quantifying randomness with infinite dimensional coherence}\label{sec:infinitecoherence}
%The randomness associated with a source is commonly quantified by the Shannon entropy function, which is well defined for all discrete random variables. Randomness is often characterized by the min entropy functions.
%Here, we first establish a connection between these two measures.
Now we consider the randomness origin and quantification in the shot-noise driven QRNG based on coherent detection. Figure.~\ref{fig:settings}(d) schematically shows its mechanism, including a quantum phase performing Fock basis measurement on certain input states and a classical phase performing a post-processing on the Fock basis measurement outcomes. Same as finite dimensional case, the true randomness origins from the Fock basis measurement breaking the infinite dimensional coherence, which cannot be directly detected as raw data since the classical noises dominates. The post-processing, subtracting the two measurement results, is able to mitigate the classical noises and let the proportion of quantum signals high enough to be detected.

We begin the randomness quantification with the quantum version of min entropy function and show that, in the asymptotic limit, when an experiment is repeated infinitely many times, the average randomness per round approaches the Shannon entropy function.
%The randomness in the shot-noise driven QRNG origins from the breaking of infinite dimensional coherence.% When quantifying the randomness of measuring pure states, one should apply (smooth) min entropy for one-shot cases and Shannon entropy for asymptotic cases.
Consider an arbitrary state $\rho_A$, after a projective measurement $\ket{i}\bra{i}$ on A, $\rho_A$ is dephased to $\rho_A^\prime=\sum_i p_i \ket{i}\bra{i}$ in the measurement basis. In the worst case, the adversary is access to the most side information of the measurement outcomes by holding a purification of $\rho_{AE}=\ket{\Psi}_{AE}\bra{\Psi}_{AE}$. And the state after measurement is $\rho_{A^\prime E}=\sum_i p_i\ket{i}\bra{i} \otimes \rho_E^i$. The one-shot randomness in the measurement outcome against such an adversary is given by conditional min-entropy \cite{tomamichel2012framework}
\begin{equation}\label{eq:quantumminentropy}
S_{min}(A^\prime|E)=\max_{\sigma_E}\sup\{\lambda\in \mathcal{R}:\rho_{A^\prime E}\leq2^{-\lambda}I_A\otimes \sigma_E\},
\end{equation}
where the dimension of $\sigma_E$ is not higher than that of $\rho_E=\mathrm{tr}_E(\rho_{AE})$. In Appendix \ref{App:minentropyreduction}, we prove that when $\rho_A$ is pure, this formula will reduce to the classical min-entropy function $H_{min}=-\log_2(\max_i p_i)$. The $\epsilon$-smooth version of Eq.~\eqref{eq:quantumminentropy}, removing extreme events, is also a one-shot randomness quantifier,
\begin{equation}
S^\epsilon_{min}(A^\prime|E)=\max_{\tilde{\rho_{AE}}}S_{min}(A^\prime|E)
\end{equation}
satisfying $\sqrt{1-F^2(\tilde{\rho}_{AE},\rho_{AE})}\leq \epsilon$, where fidelity function is defined as $F(\tilde{\rho}_{AE},\rho_{AE})=\mathrm{tr}(\tilde{\rho}_{AE} \rho_{AE})$.

\begin{figure}[htb]
\centering
\includegraphics[width=8.5cm]{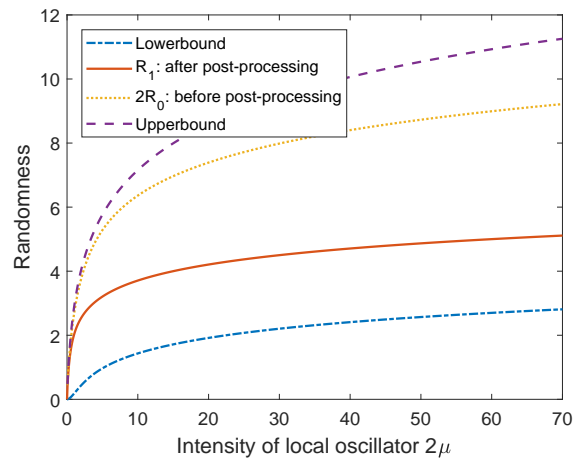}
\caption{Dependence of the randomness before and after the post-processing of subtraction on the intensity of LO. In the legend ``$R_1$: after post-processing'' refers to Eq.~\eqref{eq:idealrandomnessrate} and ``$2R_0$: before post-processing'' refers to Eq.~\eqref{eq:bilateralentropy}. The dashed line and dot-dashed line are practical upper and lower bound of randomness per sample given in Sec.~\ref{Sec:practicalbounds}, respectively.
}
\label{fig:randomnessvsphoton2}
\end{figure}

If the measurement is conducted $n$ times, in an independent and identical way, then the outputs are also independent and identically distributed (i.i.d) variables whose randomness is given by $S^\epsilon_{min}(A^{\prime n}|E^n)$. In the limit of $n\rightarrow \infty$, for any $0<\epsilon<1$
\begin{equation}\label{eq:equipartition}
\lim_{n\rightarrow \infty}\frac{1}{n}S^\epsilon_{min}(A^{\prime n}|E^n)=S(A^\prime|E)=S(A^\prime)-S(A)
\end{equation}
%\begin{equation}\label{eq:equipartition}
%\begin{aligned}
%&\lim_{n\rightarrow \infty}\frac{1}{n}S^\epsilon_{min}(A^{\prime n}|E^n)=S(A^\prime|E) \\
%&=S(A^\prime E)-S(E) \\
%&=H(\{p_i\})+\sum_i p_i S(\rho_E^i)-S(A) \\
%&=S(A^\prime)-S(A)
%\end{aligned}
%\end{equation}
where the first equation is the asymptotic equipartition property \cite{tomamichel2012framework}, the second equation is referred to Ref.~\cite{Yuan16} for finite dimensional cases, but it still holds for infinite dimensional cases since the relative entropy of coherence is a well-defined coherence measure for infinite dimensional states \cite{zhang2016quantifying}. Therefore, we can conclude that the randomness after the Fock basis measurement can be quantified with relative entropy of coherence. Fortunately, in our shot-noise driven QRNG, the state $\rho_A$ is a pure coherent state, the relative entropy of coherence reduce to Shannon entropy of the probability distribution of the measurement results,
\begin{equation}\label{eq:bilateralentropy}
\begin{aligned}
R_0&=C(\rho_A)=H(\{p^{\mathcal{P}}_j(\mu)\})\\
&=-\sum_{j=-\infty}^{\infty}p^{\mathcal{P}}_j(\mu)\log_2{p^{\mathcal{P}}_j(\mu)},
\end{aligned}
\end{equation}
where $p^{\mathcal{P}}_j(\mu)$ is given by Eq.~\eqref{eq:poisson}. After the post-processing of subtraction, the final randomness becomes
\begin{equation}\label{eq:idealrandomnessrate}
\begin{aligned}
R_1&=H(\{p^{\mathcal{S}}_j(\mu)\}) \\
&=-\sum_{j=-\infty}^{\infty}p^{\mathcal{S}}_j(\mu)\log_2{p^{\mathcal{S}}_j(\mu)},
\end{aligned}
\end{equation}
which is less than the total randomness $2R_0$ and $p^{\mathcal{S}}_j(\mu)$ is given by Eq.~\eqref{eq:skellam}. We compare the randomness before and after the subtraction, i.e., $2R_0$ and $R_1$ respectively, with respect to the intensity of the local oscillator in Fig.~\ref{fig:randomnessvsphoton2}.

%the third equation is according to the von Neumann entropy of classical-quantum state, and $S(A)=S(E)$ since $\rho_{AE}$ is a pure state, and the forth equation is because $\rho_E^i$ is a pure state and $H(\{p_i\})=S(A^\prime)$. Notice that the final result is just the relative entropy of coherence in the measurement basis.

%Figure~\ref{fig:randomnessvsphoton} shows the dependence of $R_0$ on the intensity of the local oscillator, $2\mu$, in photon count. For small values of $\mu$, the randomness is low but it sharply increases with the increase in intensity. For $2\mu > 10$, the rate increases slowly with the rise in the intensity. In this figure, we also compare $R_0$ with the bilateral entropy, which is given by the sum of the entrpy in $N_0$ and $N_1$ before their subtraction
%For varying intensity of the LO $2\mu$, the randomness changes as shown in Fig.~\ref{fig:randomnessvsphoton}. Compared with a logarithmic function $\log_2{(2\mu+1)}$, the randomness will become lower, and grow slower than the logarithmic function when $\mu$ is large enough, which is consistent with our intuition that the randomness will be upper bounded by a uniform distribution of $\{0,1,2,\cdots, 2\mu\}$ (We explain it in detail in Sec.~\ref{Sec:practicalbounds}). We also compare the final randomness with the total entropies before the subtraction, which is called bilateral entropy given by

%Our approach above is applicable if the local oscillator is an arbitrary state. The photon number statistics will be different in that case. However, the randomness origin will still depend on coherence consumption on the Fock basis {\color{red} Not sure of statement here}.

\section{Practical bounds for realistic implementations}\label{Sec:practicalbounds}
In the last section, we obtained the random number generation rate for the shot-noise driven QRNG in Eq.~\eqref{eq:idealrandomnessrate}. In this section, we try to find an upper bound $R^U$ and a lower bound $R^L$ on Eq.~\eqref{eq:idealrandomnessrate} for practical cases. The upper bound $R^U$ provides a limit on the output randomness per sample, while the lower bound $R^L$ is just randomness quantification in previous works \cite{Gabriel10,STZ10}. In the following analysis, we model some experimental parameters relevant to realistic setups. In what follows, the power of the local oscillator, which is assumed to be generated by a continuous-wave laser, by $P$, central and max frequency of the laser by $\nu$ and $\nu_m$, the response time of the photodetectors by $\tau$, the sampling frequency of ADC by $f$, and the quantization interval of ADC by $a$.
\subsection{Upper bound}
%First we prove the photon number channel model in quantum key distribution is valid for our LO source. Suppose a total photon number measurement is performed on the two output modes of beam splitter, the output of the homodyne detection will not change since the fock basis measurements commute with each other. Such a total photon number measurement also commutes with the unitary transformation by the beam splitter since the conservation of the total photon number.
The photon number of the LO within the response time follows Poisson distribution $p^{\mathcal{P}}_j(2\mu)$, where $2\mu=P\tau/(h\nu)$ is the mean photon number. The total randomness comes from two aspects, the randomness in the detection outcomes and the randomness in the Poisson distribution, i.e., $H(AB)=H(A|B)+H(B)$, where $A$ and $B$ stand for the two aspects above respectively. The maximum possible randomness for $n$-photon input is that the photon number difference $\{-n,-n+2,\cdots,n-2,n\}$ follows a uniform distribution, which corresponds to the max-entropy $\log_2(n+1)$. Then $H(A|B)<\sum_n p_n\log_2(n+1)\leq \log_2[(\sum_n p_n n)+1]$, where the second inequality is due to the concavity of logarithm function. Therefore we obtain an upper bound of randomness per sample,
\begin{equation}\label{eq:upperbound}
R^U=\log_2\left(2\mu+1\right)+H(\{p^{\mathcal{P}}_j(2\mu)\}),
\end{equation}

The upper bound of randomness generation rate is proportional to $R^U$, while the sampling frequency is constrained by the response time and Nyquist-Shannon sampling theorem \cite{nyquist1928certain,shannon1949communication}.
When the sampling frequency exceeds $1/\tau$ or $2\nu_m$, the information becomes redundant due to high autocorrelation. Therefore, the upper bound of randomness generation rate is given by
\begin{equation}\label{eq:upperbound1}
R_{\rm tot}^{\rm (max)}=\min\{\frac{1}{\tau},2\nu_m\}R^U.
\end{equation}
Considering some specific photodetectors, we list the corresponding upper bounds in Table.~

\subsection{Lower bound}
In order to find a lower bound on $R_1$, we can use the relationship $H(\{p^{\mathcal{S}}_j(\mu)\}) \geq H_{min}(\{p^{\mathcal{S}}_j(\mu)\}) = - \log_2(p^{\mathcal{S}}_0(\mu))$. However, in practice, instead of measuring $N_d$ directly, we typically measure $k N_d$, which represents the voltage/current corresponding to the photon count, where $k$ is a proportionality factor. We also need to account for the effect of quantization in the employed ADC that follows the homodyne receiver. For an ADC with a quantization interval $a$, we can only tell if the output voltage/current lies in a certain interval with width $a$. The probability, $P_J$, that the corresponding output voltage/current to the homodyne receiver will lie in the interval $[J,J+a]$ is given by
\begin{equation}
P_J=\sum_{\lceil{J/k}{}\rceil \leq j \leq \lfloor{(J+a)/k}{}\rfloor}p^{\mathcal{S}}_j(\mu).
\end{equation}
Considering the symmetric form of the Skellam distribution, shown in Fig.~\ref{fig:ADCprob}, we can then show that the min entropy for the ADC output is given by $-\log_2(P_J)$ at $J= -a/2$. Given that, at $J= -a/2$, $P_J \geq p_0$, the lower bound on $R_0$ is given by
\begin{equation}\label{eq:lowerbound}
R^L=-\log_2{\sum_{\lceil{-a/(2k)}{}\rceil \leq j \leq \lfloor{a/(2k)}{}\rfloor} p^{\mathcal{S}}_j}(\mu).
\end{equation}
%In experiments, the randomness per sample is calculated with min-entropy as a lower bound. The photodetector transforms photon intensity to current linearly with a coefficient $k$, and the ADC transforms current to voltage linearly again. We set the coefficient of the total process that transforms photon intensity to voltage to be $k'$. Given an ADC with resolution $a$, a certain voltage interval is $[J,J+a]$. When the detection result falls in this interval, a corresponding sequence of random numbers is generated. The probability of this sequence is to sum over the probabilities of events falling in $[J,J+a]$, $P_J=\sum_{J/k'<j<(J+a)/k'}p_j$. And the maximum probability is to maximize the position of the interval $P_{J,max}=\max_{J}P_J$. In our case, for a symmetric Skellam distribution, $P_J$ gets its maximum at $J=-a/2$ as shown in Fig.~\ref{fig:ADCprob}, i.e., $P_{J,max}=\sum_{-a/(2k')<j<a/(2k')}p_j$. It is straightfoward that $H_{min}(\{p_j\})>H_{min}(\{P_J\})=-\log_2{P_{J,max}}$. Then the lower bound $R^L$ is given by
Similarly, the lower bound on the total random number generation rate is given by
\begin{equation}\label{eq:randomnessexp}
R_{\rm tot}^{\rm(min)}=\min\{\frac{1}{\tau},2\nu_m\} R^L
\end{equation}
%where $2\mu= P \tau /(h\nu)$ depends on the response time of the photodetector and the intensity of the local oscillator.

\begin{figure}[htbp]
\centering
\includegraphics[width=8.5 cm]{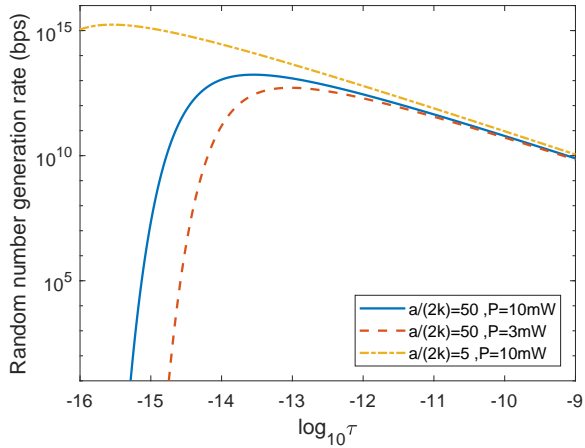}
\caption{The lower bound of random number generation rate with different resolutions of the ADC and different local oscillator intensities.}
\label{fig:randomnessvsphoton}
\end{figure}

%\subsection{Simulation for practical case}
Such a lower bound is actually applied as the randomness generation rate in experiment since it is easy to calculate. We make a comparison between the randomness upper bound Eq.~\eqref{eq:upperbound}, the lower bound Eq.~\eqref{eq:lowerbound}, and the actual randomness Eq.~\eqref{eq:idealrandomnessrate} in Fig.~\ref{fig:randomnessvsphoton2} with the ADC resolution $a/k=1$. We further simulate the randomness generation rate lower bound based on Eq.~\eqref{eq:randomnessexp} for different resolutions of the ADC and different local oscillator intensities in Fig.~\ref{fig:randomnessvsphoton}. Here we neglect the constraint of Nyquist-Shannon sampling theorem and assume the optimal sampling frequency is equal to reciprocal value of the response time of the photo detector $1/\tau$. The simulation result shows the lower bound of random number generation rate has a peak value and becomes convergent when the sampling frequency goes to infinity. This is because when $\tau \rightarrow 0$, the variance per sample also goes to zero, and the measurement result will always fall in a certain interval of the ADC, which leads to a fixed sequence with a min-entropy of zero. For practical photodetectors, the response time is at the order of $10^{-10}$ s which is much larger than the optimal value. Therefore the sampling frequency can be increased to $1/\tau \sim 10^{10}$ Hz in practical implementations.

\section{Conclusion and Outlook}
\label{sec:conc}
%{\color{red} I suggest that we rewrite this section when the results are final. Perhaps, Hongyi, you can write the frist draf summarising the results of the paper.}
In this work, we investigate the randomness quantification in shot-noise driven QRNG based on coherent detection. By characterizing the local oscillator in a quantum way, we find the outcome of homodyne detection is actually an infinite dimensional discrete variable rather than a continuous one, whose randomness is quantified by infinite-dimensional coherence. Considering experimental parameters, we calculate practical upper and lower bounds of the randomness generation rate.

As a beginning, our work provides a new point of view on the coherent detection. For future work, we may take more practical issues into consideration, such as electronic noises, bandwidth of photodetectors, and more important, intensity fluctuations of the local oscillator and input state. These intensity fluctuations will make the coherent states in our model become mixed, which will be exploited by the adversary to extract side information. The randomness quantification in this case is quite challenging.

Moreover, our technique for randomness quantification in coherent detection is applicable to other scenarios that a similar setup is used. One example is phase fluctuation extracting randomness from spontaneous emission. The bottleneck lies in how to characterize the entropy source, i.e., a coherent light carrying a random phase introduced by spontaneous emission.

Another key example is the continuous-variable quantum key distribution (CV-QKD) systems where a Gaussian-modulated coherent state by Alice is measured by a homodyne receiver at Bob's end \cite{GG02}. The key rate calculations will then involve estimating the mutual information between Alice and Bob and upper bounding the Holevo information between Alice/Bob and Eve (depending on whether direct/reverse reconciliation is in use) \cite{Devetak207}. The common assumption in such calculations is that of treating the local oscillator classically, or, equivalently, assuming that the local oscillator is of infinitely large intensity. If one wants to account for the effect of having a finite-power oscillator, then one can use the techniques we developed in this work, and the security analysis may fall in to the same framework of discrete variable quantum key distribution (DV-QKD).
%In this work, we investigate the basic mechanism of homodyne detection from the first principle, and apply the physical model into shot-noise QRNG scheme based on homodyne detection of a vacuum input. We show that the outcome of homodyne detection is actually a infinite dimensional discrete variable rather than a continuous one. Moreover, we interpret the randomness origin and calculate the randomness output in such a QRNG scheme with infinite dimensional coherence. Subsequently we give an upper bound of the coherence as the limit of randomness generation rate and a lower bound as the randomness certification in experiment. A simulation of the shot-noise QRNG scheme shows that the random number generation rate has a peak value of $10^{13}$ bps. Finally we show a possible alternative approach for the security proof of continuous variable quantum key distribution.
%Moreover, for a photodetector, there will be a mapping from the photon number to current. Suppose

\acknowledgments
The authors acknowledge J.~Ma, M. Plenio and X.~Yuan for the insightful discussions. This work is supported by National Key R\&D Program of China (2017YFA0303900, 2017YFA0304004), the National Natural Science Foundation of China Grant No.~11674193, and the UK EPSRC Grant EP/M013472/1. All data generated in this paper can be reproduced by the provided methodology and equations.

\appendix
\section{Classical model of homodyne detection} \label{App:ClassicalModel}
Homodyne detection settings, made up of a beam splitter and two photodetectors, have two inputs: a local oscillator (LO), which can be described as a strong coherent state $\ket{\alpha_{LO}}$, and a signal state $\rho$. After a transformation from photon intensity to current by the photodetector, a subtraction of current is performed to mitigate the classical electronic noise.

The output in the homodyne detection is given by the operator,
\begin{equation}
\delta\hat{i}=\hat{i}_1-\hat{i}_2=k(\hat{a}^\dag_{LO}\hat{a}+\hat{a}^{\dag}\hat{a}_{LO}).
\end{equation}
where $\hat{a}$ and $\hat{a}_{LO}$ are annihilation operators of the input optical modes. And for a photodetector, we make an assumption that the current is proportional to photon number and the coefficient is $k$.
Note that all the calculation above is in the Heisenberg picture. Hence the expectation value of $\delta\hat{i}$ is $Tr(\delta \hat{i}\rho\otimes \ket{\alpha_{LO}}\bra{\alpha_{LO}})$ and the variance is $Tr(\delta \hat{i}^2\rho\otimes \ket{\alpha_{LO}}\bra{\alpha_{LO}})-(Tr(\delta \hat{i}\rho\otimes \ket{\alpha_{LO}}\bra{\alpha_{LO}}))^2$. When the intensity of the LO is strong enough, the homodyne detection can be regarded as a measurement of quadratures as an approximation.
Considering the phase of the LO, $\alpha_{LO}=|\alpha_{LO}|e^{i\phi}$, the expectation and variance can be rewritten as
\begin{equation}\label{eq:expectationandvariance}
\begin{aligned}
& \langle \delta\hat{i} \rangle =2k|\alpha_{LO}|\mathrm{tr}(\hat{x}(\phi)\rho) \\
& \langle \delta\hat{i}^2 \rangle-\langle \delta\hat{i}\rangle^2 \\
&=4k^2|\alpha_{LO}|^2[\mathrm{tr}(\hat{x}^2(\phi)\rho)-(\mathrm{tr}(\hat{x}(\phi)\rho))^2]+k^2\mathrm{tr}(\hat{a}^\dag \hat{a}\rho),
\end{aligned}
\end{equation}
where $\hat{x}(\phi)=(\hat{a}e^{-i\phi}+\hat{a}^{\dag}e^{i\phi})/2$ is a quadrature of the signal state depending on $\phi$. When $\phi=0$ or $\pi/2$, it corresponds to $\hat{x}=(\hat{a}+\hat{a}^\dag)/2$ or $\hat{p}=(\hat{a}-\hat{a}^\dag)/2$ quadrature respectively, that is, the quantity measured by the homodyne detection depends on the phase $\phi$ of the LO.

\section{Quantum min-entropy can reduce to its classical counterpart} \label{App:minentropyreduction}
In this section we prove the quantum min-entropy will reduce to the classical min-entropy function when the adversary has no side information of the measurement outcomes, i.e, $\rho_A$ is a pure state. We begin with Eq.~\eqref{eq:quantumminentropy}
\begin{equation}\label{eq:quantumtoclassicalminentropy}
\begin{aligned}
S_{min}(A^\prime|E)&=\max_{\sigma_E}\sup\{\lambda\in \mathcal{R}:\rho_{A^\prime E}\leq2^{-\lambda}I_A\otimes \sigma_E\}	\\
&=\min_{\sigma_E}\inf\{p \in \mathcal{R}:pI_A \otimes \sigma_E \geq \rho_{A^\prime E}\} \\
&=\min_{\sigma_E}\inf\{p \in \mathcal{R}:pI_A \otimes \sigma_E \geq \rho_{A^\prime}\otimes \rho_E\} \\
\end{aligned}
\end{equation}
where $p=2^{-\lambda}$, the last equation is because $\rho_A$ is pure and after the measurement on $A$, $\rho_{A^\prime E}$ is also a product state. Now we need to let $p$ as small as possible such that $pI_A \otimes \sigma_E \geq \rho_{A^\prime}\otimes \rho_E$ which can be rewritten as
\begin{equation}\label{eq:appendixtarget}
\sum_i \ket{i}\bra{i}\otimes (p\sigma_E-p_i \rho_E)	
\end{equation}
We only need to consider $p\sigma_E-p_i \rho_E\geq 0$. Note that $\rho_E$ is a pure state with only one non-zero eigenvalue $\eta=1$ in its spectrum. In order to let $p$ as small as possible, the best choice is to let $\sigma_E$ also be a pure state $\sigma_E=\rho_E$ and $p\geq p_i$. Consider all decomposition components in Eq.~\eqref{eq:appendixtarget}, $p=\max_i p_i$ and $\lambda=-\log_2(\max_i p_i)$ which is just the classical min-entropy function.
%\end{appendix}

%\section*{Supplemental Documents}
%See Supplement 1 for supporting content.

%\bigskip \noindent See \href{link}{Supplement 1} for supporting content.

% Bibliography
%\bibliographystyle{apsrev4-1}
\bibliography{bibvacuum}

\end{document}